\documentclass[12pt,3p,twocolumn]{elsarticle}
\pdfoutput=1
\usepackage{amssymb}
\usepackage{amsmath}
\usepackage{algorithm}
\usepackage{algorithmic}
\usepackage{subfigure} 
\usepackage{url}
\usepackage{color}

\biboptions{sort}

\begin{document}

\begin{frontmatter}

\title{A Second-Order Distributed Trotter-Suzuki Solver with a Hybrid Kernel}


\author{Peter Wittek\corref{cor1}}
\ead{peterwittek@acm.org}
\cortext[cor1]{Swedish School of Library and Information Science, University of Bor\aa s, Allegatan 1, Bor\aa s, S-501 90, Sweden}
\author{Fernando M. Cucchietti\corref{cor2}}
\ead{fernando.cucchietti@bsc.es}
\cortext[cor2]{Barcelona Supercomputing Center (BSC-CNS), Edificio NEXUS I, Campus Nord UPC, Gran Capit\'{a}n 2-4, 08034 Barcelona, Spain}

\begin{abstract}
The Trotter-Suzuki approximation leads to an efficient algorithm for solving the time-dependent Schr\"odinger equation.
Using existing highly optimized CPU and GPU kernels, we developed a distributed version of the algorithm that runs efficiently on a cluster. Our implementation also improves single node performance, and is able to use multiple GPUs within a node. The scaling is close to linear using the CPU kernels, whereas the efficiency of GPU kernels improve with larger matrices. We also introduce a hybrid kernel that simultaneously uses multicore CPUs and GPUs in a distributed system. This kernel is shown to be efficient when the matrix size would not fit in the GPU memory. Larger quantum systems scale especially well with a high number nodes. The code is available under an open source license.
\end{abstract}


\end{frontmatter}

\section{Introduction}
The evolution of a general quantum system is described by the time-dependent Schr\"odinger equation. 
The generic solution of this equation involves calculating a matrix exponential, which is formally simple.
However, computer implementations must consider several factors to achieve high performance and high accuracy -- usually making a trade-off between these two indicators.

There is a wide range of numerical approaches to calculating a matrix exponential. However, since they are approximate, not all may preserve some desired analytical property of the original matrices. 
This is crucial for example with the quantum evolution operator -- the exponential of the Hamiltonian matrix appearing in the Schr\"odinger equation, -- which must be unitary in order to conserve the total probability. The reference method for calculating a matrix exponential is to diagonalize the matrix using an eigendecomposition, which is typically computationally intensive. While efficient eigendecomposition algorithms exist that use multicore CPU and multiple GPUs in a system \cite{tomov2012magma}, distributed variants that use several computer nodes in a cluster are hard to parallelize with close to ideal efficiency \cite{ozdogan2007scaling}.
Traditional numerical integrators like the Runge-Kutta algorithm do not conserve unitarity, and unitary algorithms like the Crank-Nicholson scheme involve inverting a large matrix.

The Trotter-Suzuki algorithm approaches the problem through a slightly different angle. It decomposes the Hamiltonian as a sum of matrices that are easy to exponentiate \cite{de1996computer}, which are then used to approximate the exponential of the full Hamiltonian. The end result is an algorithm that is easy to parallelize. For the case of a single particle in real space that we treat here, the algorithm discretizes the domain with a finite mesh and calculates the pairwise evolution between neighboring sites in the mesh. The Trotter-Suzuki algorithm has been successfully used \cite{de1996computer,cucchietti2002decoherence,deraedt2000quantum}. Efficient kernels for contemporary multicore CPUs and GPUs have already been developed \cite{bederian2011boosting}.

This paper introduces a distributed variant of the Trotter-Suzuki algorithm using existing high-performance kernels. Our implementation improves the single-node efficiency of the CPU and GPU kernels, it is able to use multiple GPUs in a single system, and also introduces a hybrid kernel to deal with large quantum systems that do not fit the GPU memory. The implementation also scales in a cluster, the CPU variant shows an almost linear speedup with the number of nodes, whereas the GPU variant is more efficient when the device has a higher load.

The rest of this paper is organized as follows. Section \ref{tsa} gives a brief overview of the foundations of the Trotter-Suzuki algorithm. Section \ref{distributed} provides the details of the distributed kernels, and Section \ref{discussion} discusses the benchmark results on a large cluster. Finally, Section \ref{future} concludes the paper and offers an insight on our future work.

\section{Trotter-Suzuki Algorithm and Efficient Kernels}\label{tsa}

The non-relativistic Schr\"odinger equation describing the evolution of a quantum system is
\begin{equation}
-\imath \hbar \frac{\partial |\psi(t)\rangle}{\partial t} = \mathcal{H}(t) |\psi(t)\rangle,
\label{schrodinger}
\end{equation}
where $\mathcal{H}(t)$ is the Hamiltonian of the system, and $|\psi(t)\rangle$ the state of the system at time $t$. By choosing a basis, the Hamiltonian can be written as a Hermitian matrix $H$.
The formal solution to the Schrodinger equation for a time-independent Hamiltonian \footnote{The general time-dependent form of the 
Trotter-Suzuki algorithm is a trivial extension of the one we present \cite{poulin2011quantum}.} is given by
\begin{equation}
|\psi(t)\rangle=e^{-\frac{\imath}{\hbar}Ht}|\psi(0)\rangle \equiv U(t) |\psi(0)\rangle,
\label{evolution}
\end{equation}
where $|\psi(0)\rangle$ is the initial state of the system, with norm $|\langle\psi(0)|\psi(0)\rangle|^2=1$, 
and $U(t)$ is the quantum evolution operator associated to Hamiltonian $H$. Since $H$ is Hermitian, it is easy to see that the evolution operator is unitary, and that
 the norm of the state vector is constant over time, $|\langle\psi(t)|\psi(t)\rangle|^2=|\langle\psi(0)|U^{\dagger}U|\psi(0)\rangle|^2=|\langle\psi(0)|\psi(0)\rangle|^2=1$
Thus, it is crucial that the numerical solution of the evolution operator be unitary \cite{de1996computer}, or the norm of the wave function --  which gives the total probability of finding the particle somewhere, and must equal one -- would not be conserved.

The Trotter-Suzuki algorithm decomposes the Hamiltonian into small diagonal or block diagonal matrices, where the exponential is easy to compute. The decomposition is based on the Trotter formula \cite{trotter1959product}:
\[
e^{x(A+B)}=\lim_{n\to\infty}\left(e^{xA/n}e^{xB/n}\right)^{n},
\]
where $A$ and $B$ are $M\times M$ matrices. For sufficiently large $n$, $e^{x(A+B)/n}\approx e^{xA/n}e^{xB/n}$. 
The Trotter formula is readily generalized to the case of more than two contributions to $H$ by writing $H = \sum_{i=1}^{p}H_{i}$. This allows for choosing a decomposition that can be exploited to construct
efficient algorithms. The error of the Trotter formula scales as $(x/n)^2$ times the norm of the commutator between $A$ and $B$. Higher order approximations were later developed by Suzuki \cite{suzuki1990fractal,suzuki1993general},
who obtained expressions that are unitary for all orders.

To explain the algorithm, let us assume a one-dimensional Hamiltonian with the form $\mathcal{H}=-\frac{\hbar^2}{2m}\frac{\partial^2}{\partial x^2}+\mathcal{V}(x)$, 
where $m$ is the mass of the particle, and $\mathcal{V}(x)$ the potential energy. 
Discretizing the Schr\"odinger equation 
using a finite mesh of points spaced by a distance $a$, the Hamiltonian matrix $H$ is like that of a tight-binding chain --
a tridiagonal matrix. Such a matrix can be split as a sum of a diagonal matrix, and two block-diagonal commuting matrices made up of $2\times 2$ matrices:
\[
H = H_{0} + H_{1} + H_{2},
\]
where the components are as follows. The diagonal matrix $H_{0}$ is written as 
\[
 H_{0} =
 \begin{pmatrix}
  \epsilon_1 & 0 & & \\
  0 & \epsilon_2 & 0 &  &  \\
    &  & \ddots &   \\
   &  & 0 & \epsilon_{L}
 \end{pmatrix},
\] 
where $\epsilon_{l}=\mathcal{V}(na)+\hbar^2/ma^2$ is the effective energy at site $l$, $L$ is the number of sites. The block diagonal matrices $H_{1}$ and $H_{2}$ are written as 
\[
 H_{1} =
 \begin{pmatrix}
  0 & V &   &   &   \\
  V & 0 & 0 &   &  & \\
    & 0 & 0 & V &  & & \\
    &   & V & 0 &\\
    &   &   &\ddots &   \\
    &   &   &   & 0 & 0
 \end{pmatrix},
\]
and
\[
 H_{2} =
 \begin{pmatrix}
  0 & 0 &   & & \\
  0 & 0 & V & & \\
    & V & 0 & & \\
    &   &   & \ddots & \\
    &   &   & & 0 & V \\
    &   &   & & V & 0
 \end{pmatrix},
\]
where $V=-\hbar^2/2ma^2$ is the tunneling element between the sites. The exponential of a block matrix is itself a block matrix build from exponentials of $2\times 2$ matrices. These plane rotation matrices can be written as 
\[
 M =
 \begin{pmatrix}
  \cos\frac{\Delta t}{\hbar}|V| & -\imath \sin\frac{\Delta t}{\hbar}|V|  \\
  -\imath \sin\frac{\Delta t}{\hbar}|V| & \cos\frac{\Delta t}{\hbar}|V|\\
 \end{pmatrix},
\]
where $\Delta t$ is the discrete time step. Using the above decomposition, the first-order approximation of the unitary time step evolution operator is given by

\[
U_{1}\left(\Delta t\right)=e^{-\imath\frac{\Delta t}{\hbar} H_{0}}e^{-\imath\frac{\Delta t}{\hbar} H_{1}}e^{-\imath\frac{\Delta t}{\hbar} H_{2}}.
\]

Approximants correct up to second order are obtained by symmetrization \cite{suzuki1985decomposition,de1996computer}:
\begin{equation}
U_{2}\left(\Delta t\right)=U^T_{1}\left(\Delta t\right)U_{1}\left(\Delta t\right),
\label{secondorder}
\end{equation}
where $A^T$ is the transpose of matrix $A$.
Extending the algorithm for more than one dimension is also straightforward, as we can simply perform a decomposition of the Hamiltonian into five parts: the diagonal energies, and two terms for each dimension.
Our implementation is based on the second order formulation of Equation \ref{secondorder}. Higher order approximants are expressed as a sequence of applications of the first and second order operators. For example, the fourth order evolution operator is
\begin{eqnarray}
U_{4}\left(\Delta t\right)&=&U_{2}\left(p\Delta t\right)U_{2}\left(p\Delta t\right) \nonumber
\\ & & U_{2}\left((1-4p)\Delta t\right) \nonumber \\
& & U_{2}\left(p\Delta t\right)U_{2}\left(p\Delta t\right),
\label{fourthorder}
\end{eqnarray}
where $p=(4-4^{1/3})^{-1}$ \cite{suzuki1990fractal}.

Because of its structure, the cost of applying any order of the Trotter-Suzuki operator scales linearly in time and memory. A general external potential is straightforward to implement and always adds a cost $L$ in time. Therefore, we perform our benchmarks assuming a flat potential landscape, $\mathcal{V}(x)=constant$, which leads to $H_0$ inducing a global phase factor in the wave function which we can ignore.

Notice in the above discussion that at no point an assumption was made about the ``importance" of a particular contribution to the Hamiltonian. This is the reason why the Trotter-Suzuki approach can be used where perturbation methods break down \cite{de1996computer}.

\section{Distributing the Workload Across a Cluster}\label{distributed}
\begin{figure}[htb!]
  \begin{center}
      \includegraphics[width=\columnwidth]{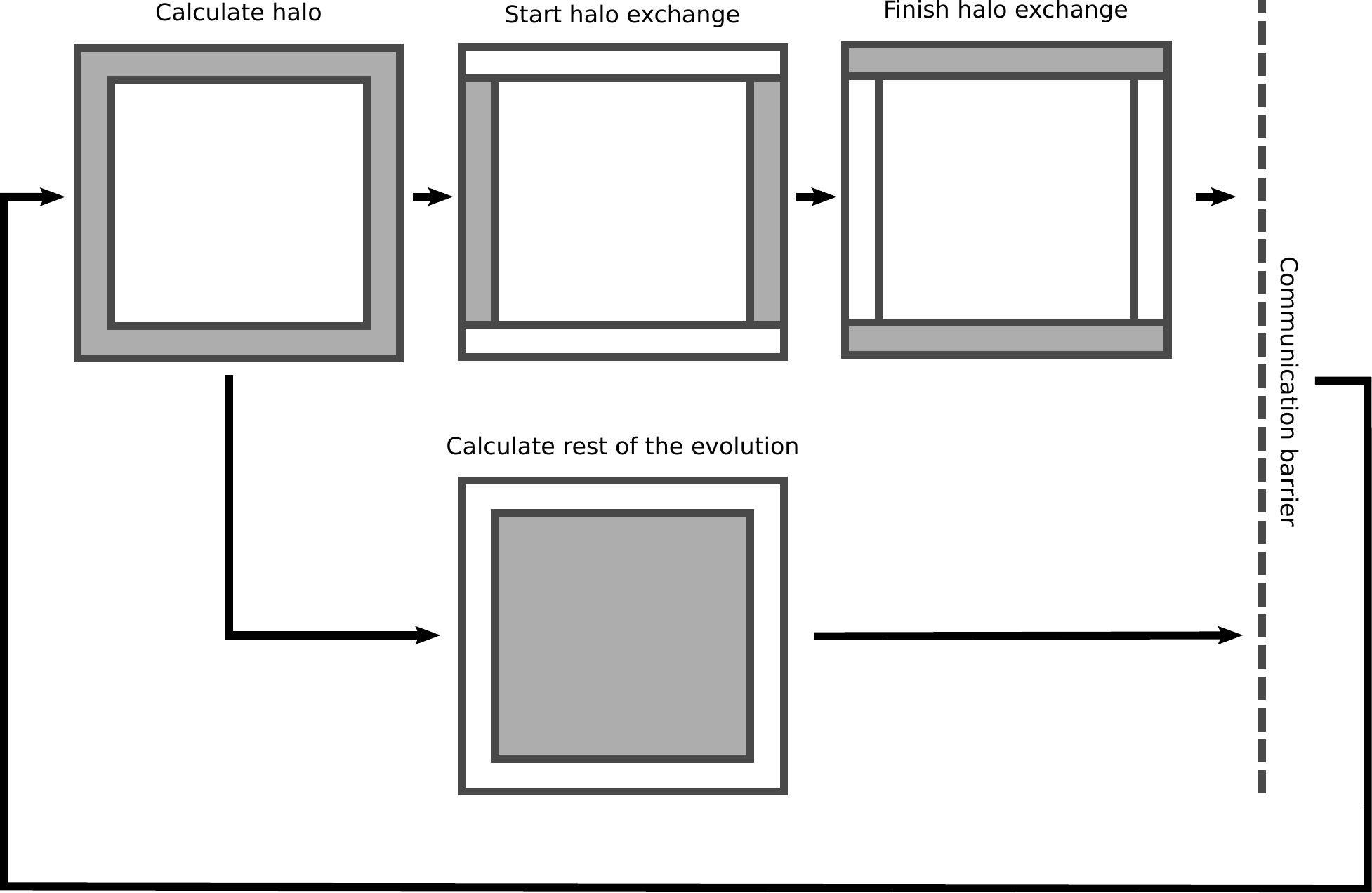}
    \caption{A general overview of communication and computation pattern within one iteration in the distributed implementation}
    \label{overview}
  \end{center}
\end{figure}

We took the optimized CPU and GPU kernels of \cite{bederian2011boosting} as our starting point. These kernels use a double-buffered data access pattern: the result is not calculated in place, but to a new buffer, and when a new iteration starts, the buffers are swapped. There are two CPU kernels: cache-optimized, and another cache-optimized that is further optimized to use the SSE instruction set of the CPU. 

Cache optimization means that the input data is divided into blocks. A similar block division is present in the GPU kernel, where, instead of a hardware cache, the shared memory of the simultaneous multiprocessors is used to fetch and explicitly cache data. This block division results in extra calculations: a halo for each block has to be computed to get the correct results for the internal cells. This extra work pays off by the benefit of cached access to the data.

The unit of calculation of our distributed version is a process. Using a similar computational structure to the above, we refer to a block assigned to a process as a tile. The block division in a single node is simple: the halo computed by a block is simply discarded, and the next block reads its own halo from the main memory. In a distributed version the halos between the tiles have to be sent across the processes. Using a two-dimensional grid of processes, a tile contains elements of halos belonging to a total of eight other tiles: left, right, top, and bottom neighbours, and also the four diagonal neighbours. To minimize the number of communication requests, a wave pattern is used in the communication: left and right neighbours receive the halo first. This halo has the height of the inner cells of the tile (see Figure \ref{overview}). Then the horizontal halos are sent to the top and bottom neighbours -- the width is the full tile width. In this way the appropriate corner elements are propagated to the diagonal neighbours.

Communication is performed asynchronously, but there is a communication barrier between the left-right and the top-bottom halo exchanges due to data dependency. The generic approach to computation and calculations attempts to overlap the two by as much as possible, calculating the halo first and starting the communication simultaneously to the calculation of the rest of the evolution step (Figure \ref{overview}).

The step of starting the halo exchange initiates the asynchronous left-right halo exchange. The last step, finishing the halo exchange, has a communication barrier waiting for the first exchange to finish. After this, it initiates the asynchronous top-bottom halo exchange, and has a second barrier, waiting for this communication to finish. There are variations to this pattern, as detailed below.

\subsection{The CPU kernels}
The cache optimized kernels of \cite{bederian2011boosting} used OpenMP, a directive-driven parallelization of the code to use the power multicore architecture \cite{dagum1998openmp}. We found that using our explicit parallelization, the overall performance was better by 30 \% if we used the same number of process within a node as there are cores. Hence, even single node execution is accelerated by our approach. This finding corresponds well with other benchmarks that compare OpenMP parallelism with more explicit forms of parallelism \cite{krawezik2003performance}.

Finishing the halo exchange cannot be efficiently overlapped with communication. This means that once the computation of the iteration is completed, there is a communication overhead while the vertical halos are exchanged. 

\subsection{The GPU kernel}
The GPU kernel had to be adjusted to work with communication. As it is customary in GPU programming, the device keeps the double-buffered matrices in its own memory, whereas MPI communication is performed from the host memory. This increases complexity as asynchronous memory copies from the device and to the device have to be performed.

To work with such transfer efficiently, we implemented the kernel with streams. A GPU stream is basically a queue of tasks for the GPU to perform: kernel execution, memory copies from and to the device. Given two streams, memory copies in one stream  can overlap with computation in the other stream. We use two streams, queueing the halo computation and the memory copies in stream one, and the computation of the rest of cells in stream two. Since a kernel launch with streams returns the control to the calling process immediately, this also means that while stream two is executing, every halo exchange can be completed before the iteration is finished. Hence the distributed GPU version has a much better communication efficiency than the CPU variant.

The communication routine typically does an internal buffering of the data to be sent. This extra memory copy is unseen by the user. To avoid it, we use pinned memory, a specific way to allocate host memory for data that will be exchanged with the GPU. Pinned memory avoids internal buffering when it comes to communication, hence further increasing communication performance.

We use one process per GPU and single-node execution with multiple GPUs is possible.

\subsection{The hybrid kernel}
Using streams means most cores of a multicore CPU idle while waiting for the GPU to complete the calculations. GPUs also have less memory, limiting the size of the quantum system for which the evolution is computed. To address these two issues, we introduce a hybrid kernel.

The algorithm first calculates the maximum amount of the tile that can be computed on the GPU assigned to the process. Then, using only one stream, it launches the kernel for the corresponding elements on the internal area of the tile. After the asynchronous kernel launches, it proceeds to calculate the halo and start the halo exchange. Once the halo exchange is initiated, the elements not in the halo and not on the GPU are computed by the CPU. Finishing the halo exchange includes an extra step, an internal halo exchange between the part of the tile associated with the GPU and the rest of the matrix.

This kernel also uses one process per GPU. This means that in an eight-core system with two GPUs, six cores would be idle. To utilize the power of the extra cores, we use the same directive-driven parallelism as the original CPU kernels of \cite{bederian2011boosting}, relying on OpenMP. Hence all cores and all GPUs contribute to the work.

\begin{figure}[ht!]
  \begin{center}
      \includegraphics[width=\columnwidth]{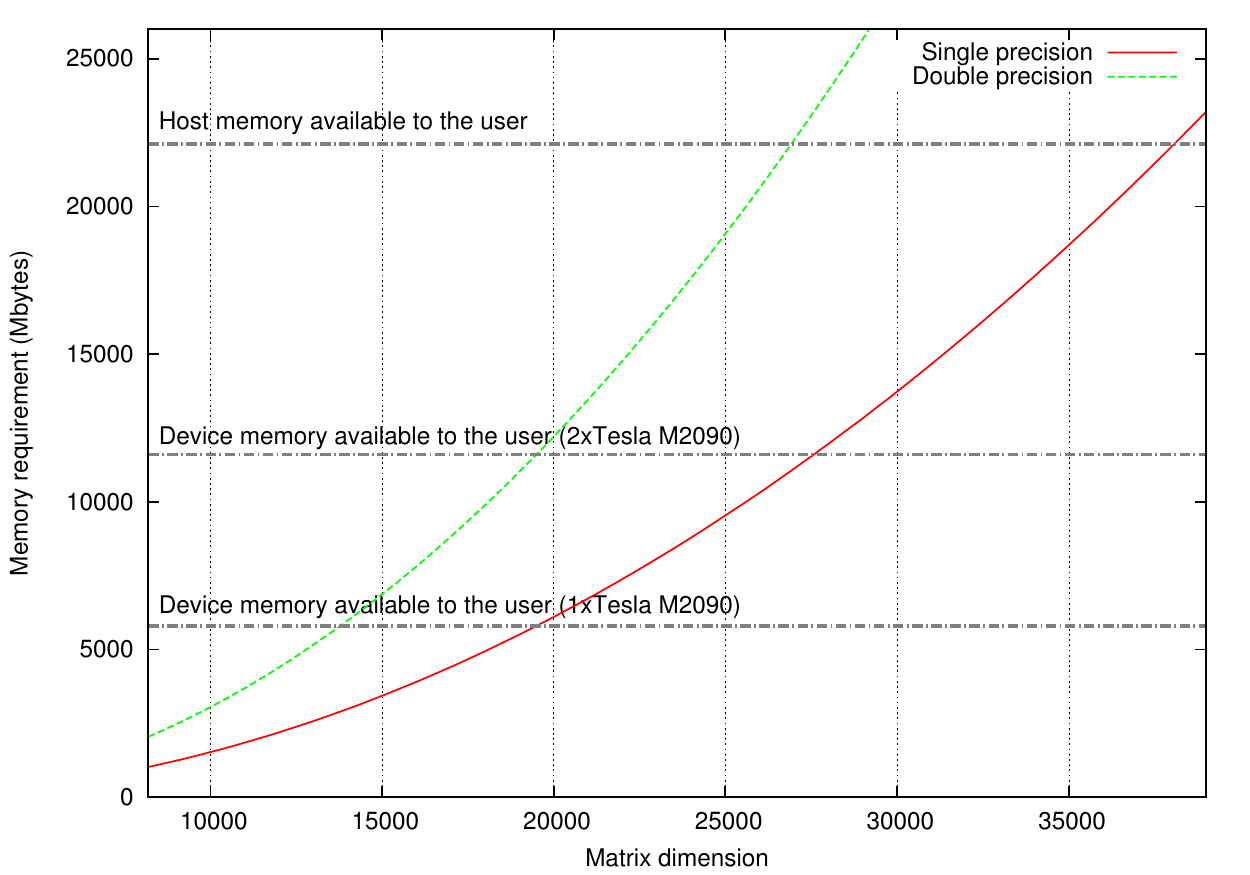}
    \caption{Memory requirements of the Trotter-Suzuki algorithm on a single node}
    \label{memory}
  \end{center}
\end{figure}

\section{Discussion}\label{discussion}

\begin{figure*}[ht!]
\centering
\subfigure[Single precision]{
\includegraphics[width=5.5cm]{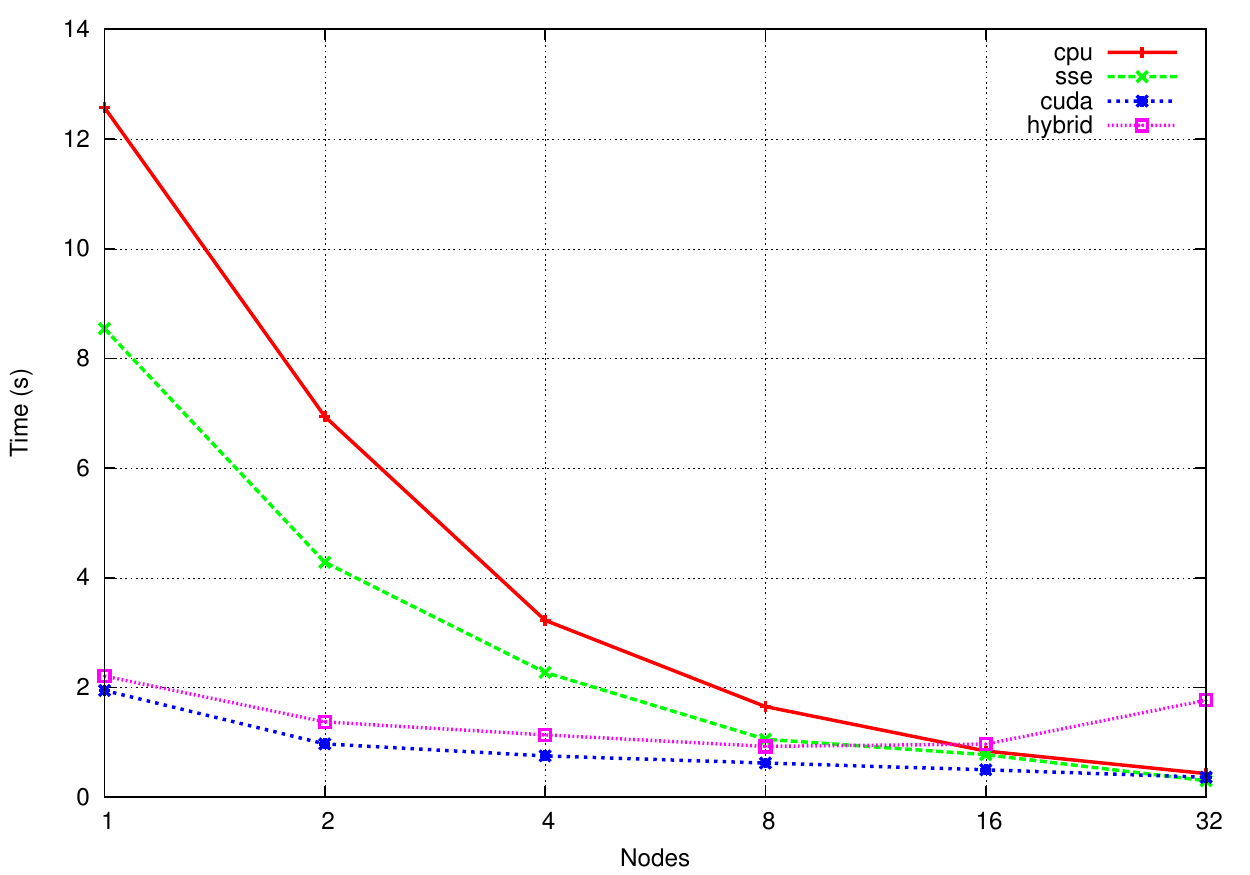}
\label{fig:subfig11}
}
\subfigure[Double precision]{
\includegraphics[width=5.5cm]{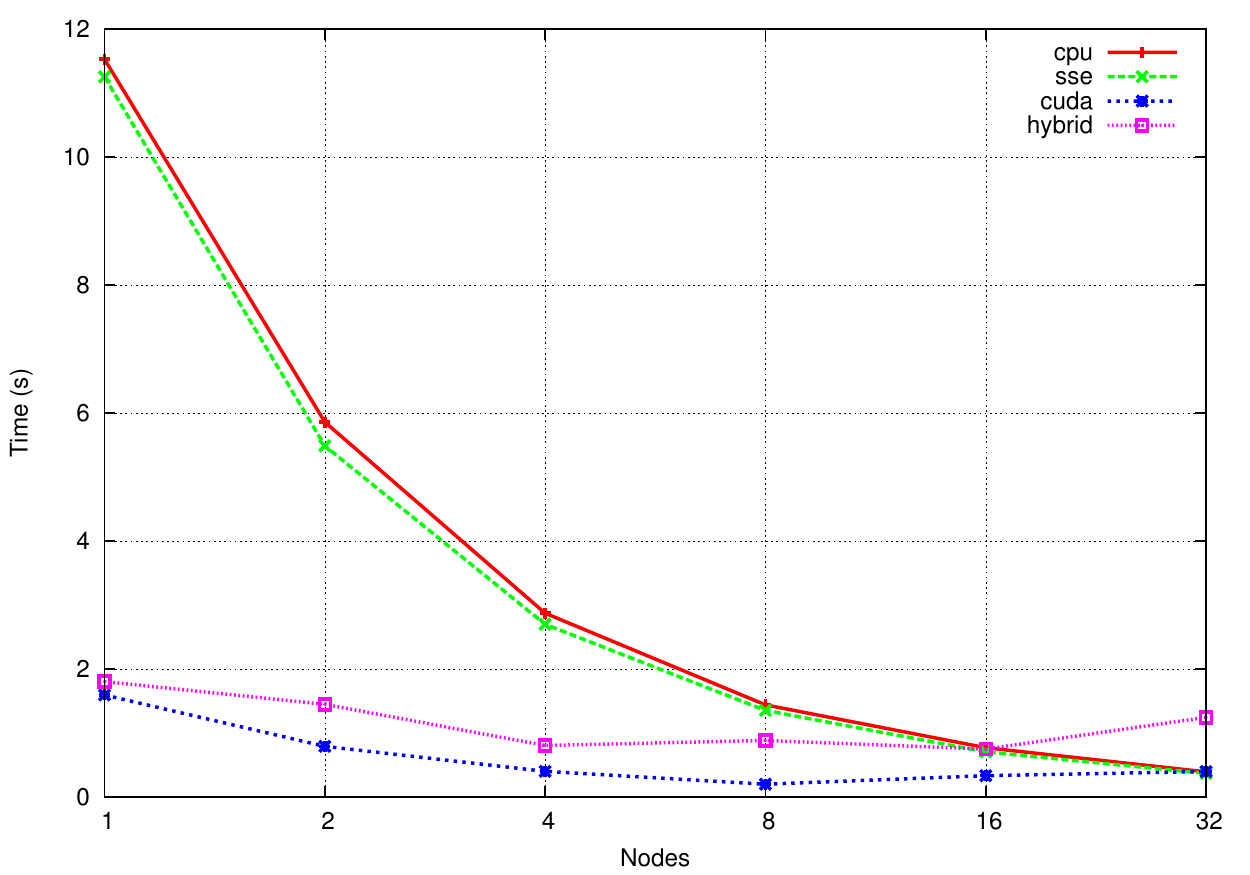}
\label{fig:subfig12}
}
\caption{Execution time, linear system size: 24576 (single precision), 17472 (double precision)}
\label{fig:exec2}
\end{figure*}

\begin{figure*}[ht!]
\centering
\subfigure[Single precision]{
\includegraphics[width=5.5cm]{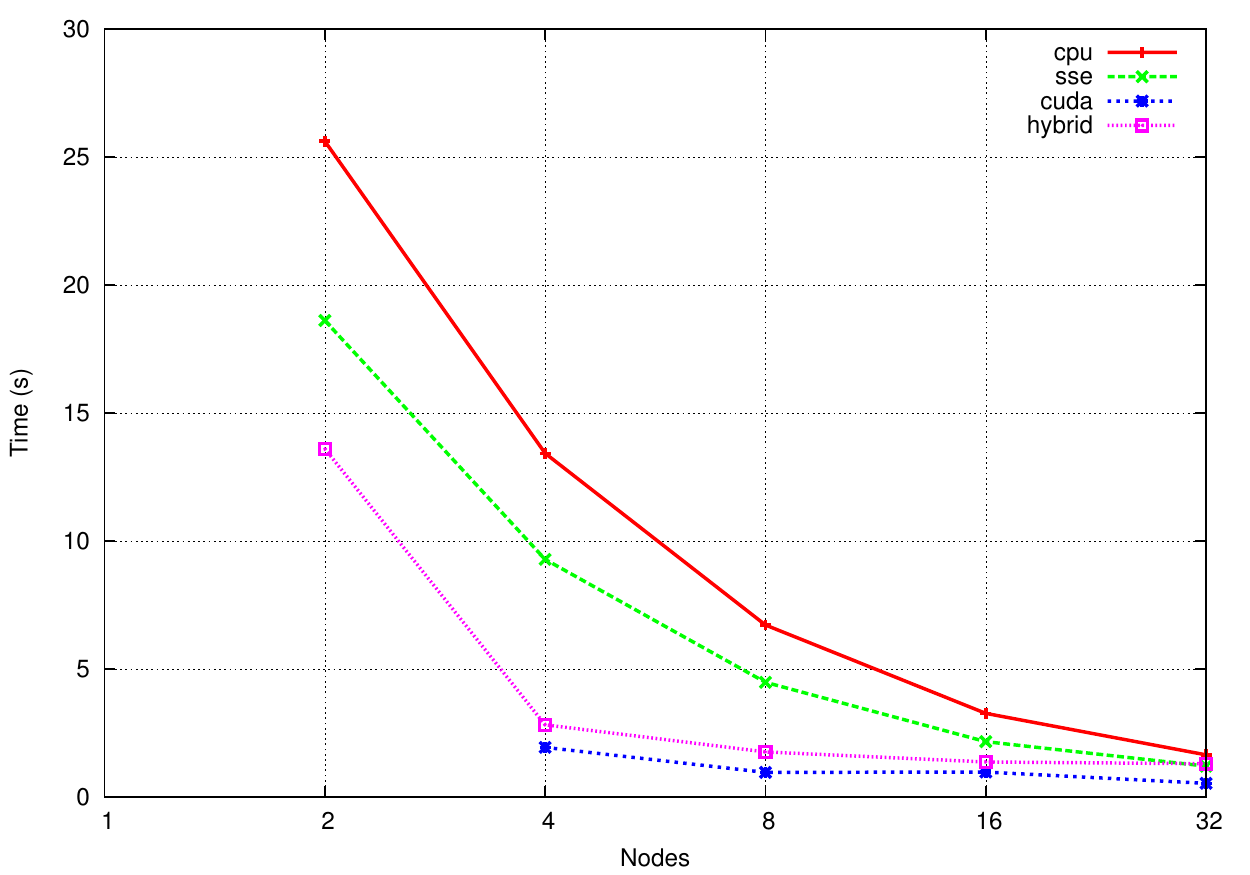}
\label{fig:subfig21}
}
\subfigure[Double precision]{
\includegraphics[width=5.5cm]{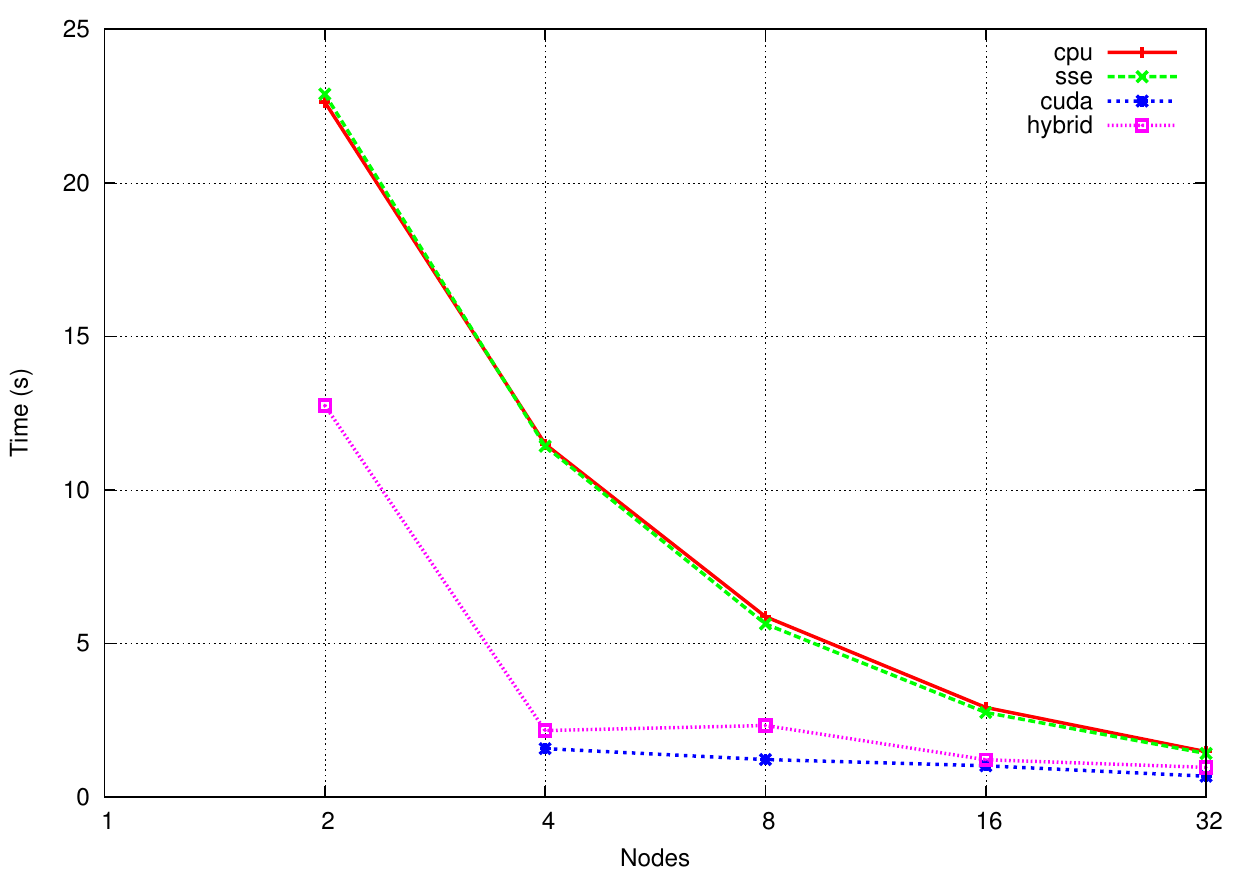}
\label{fig:subfig22}
}
\caption{Execution time, linear system size: 49152 (single precision), 34944 (double precision)}
\label{fig:exec4}
\end{figure*}

\begin{figure*}[ht!]
\centering
\subfigure[Single precision]{
\includegraphics[width=5.5cm]{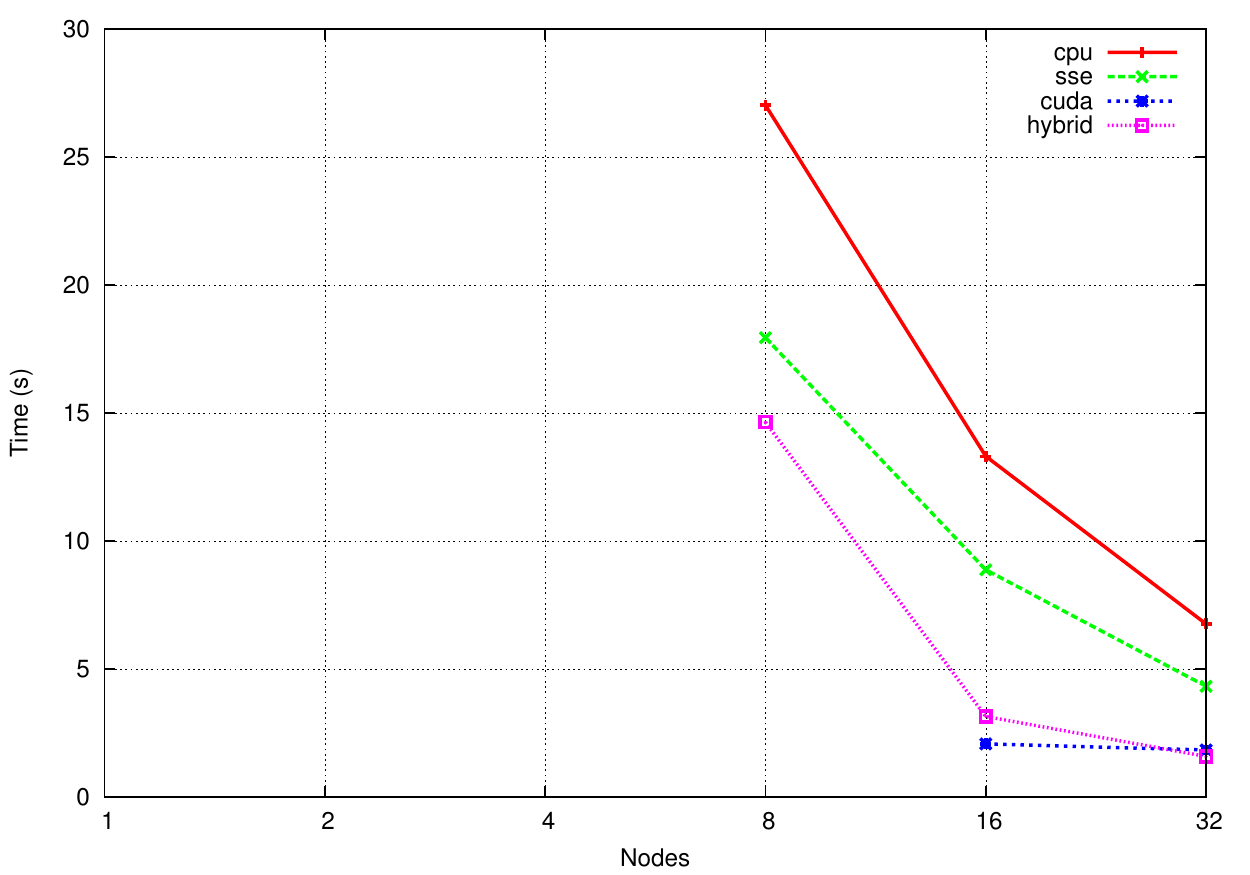}
\label{fig:subfig31}
}
\subfigure[Double precision]{
\includegraphics[width=5.5cm]{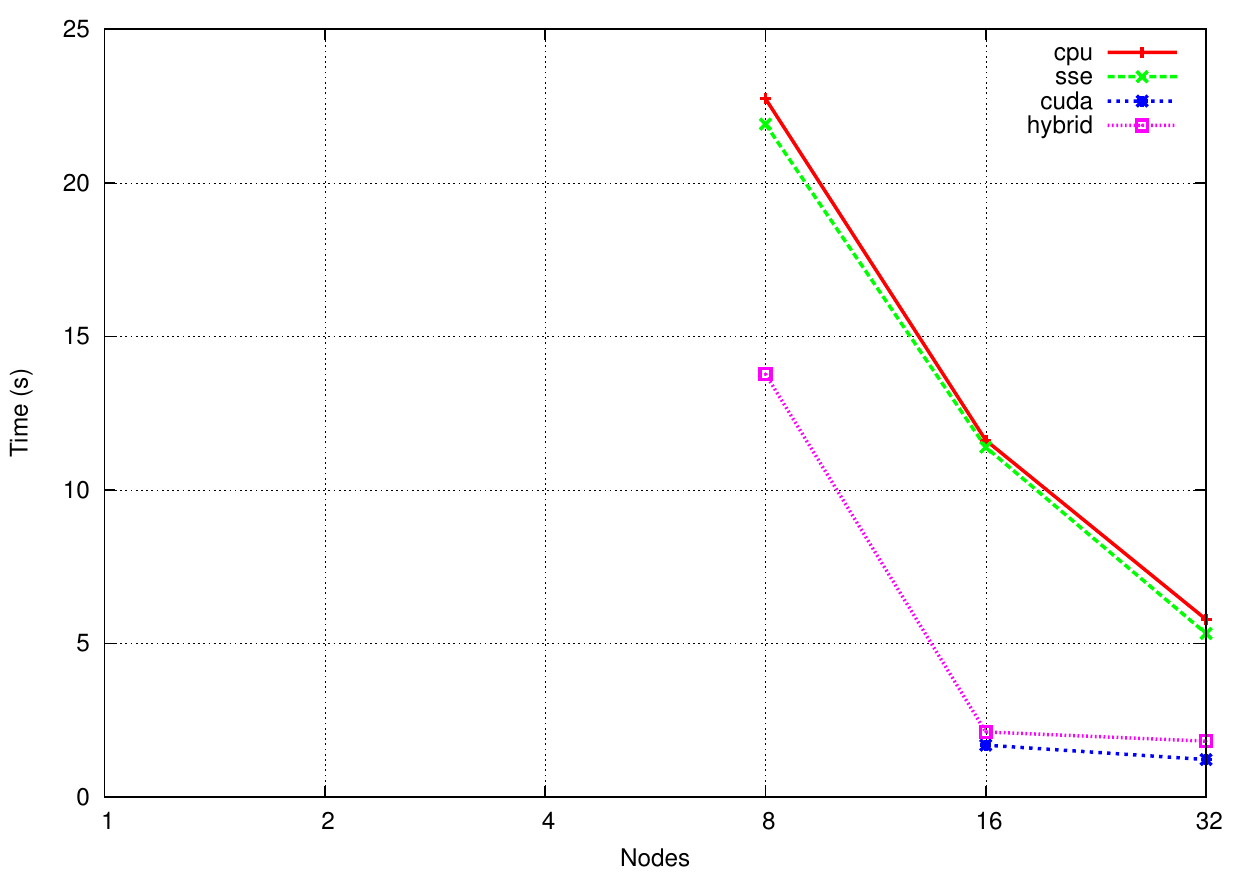}
\label{fig:subfig32}
}
\caption{Execution time, linear system size: 98304 (single precision), 69888 (double precision)}
\label{fig:exec9}
\end{figure*}

\subsection{Experimental configuration}
The implementation of the distributed algorithm used MPI for communication\footnote{The source code is available at \url{https://github.com/peterwittek/trotter-suzuki-mpi}}. We used bullx MPI, which is compatible with the MPI 2.1 standard, and it is built around OpenMPI. The compiler was the Intel Compiler Chain, with all optimization turned on and OpenMP enabled. While these are proprietary tools, the code can also be compiled with open source software, such as OpenMPI and GCC. The GPU code was implemented with CUDA and compiled with CUDA 4.0, running with the corresponding runtime.

The benchmarks were performed on the Minotauro cluster at the Barcelona Supercomputing Center. Every node has two Intel Xeon E5649 six-core processors with 12MB of cache memory, clocked at 2.53GHz, running Linux operating system with 24 GByte of RAM memory. Every node is equipped with two NVIDIA M2090 cards, each one with 512 CUDA cores and 6 GByte of GDDR5 memory. The MPI communication across the nodes is through an Infiniband Network.

Figure \ref{memory} illustrates the memory constraints on this cluster with respect to the size of the quantum system. The benchmarks below were chosen so as to maximize the memory usage on different cluster sizes.

\begin{figure*}[ht!]
\centering
\includegraphics[width=0.8\textwidth]{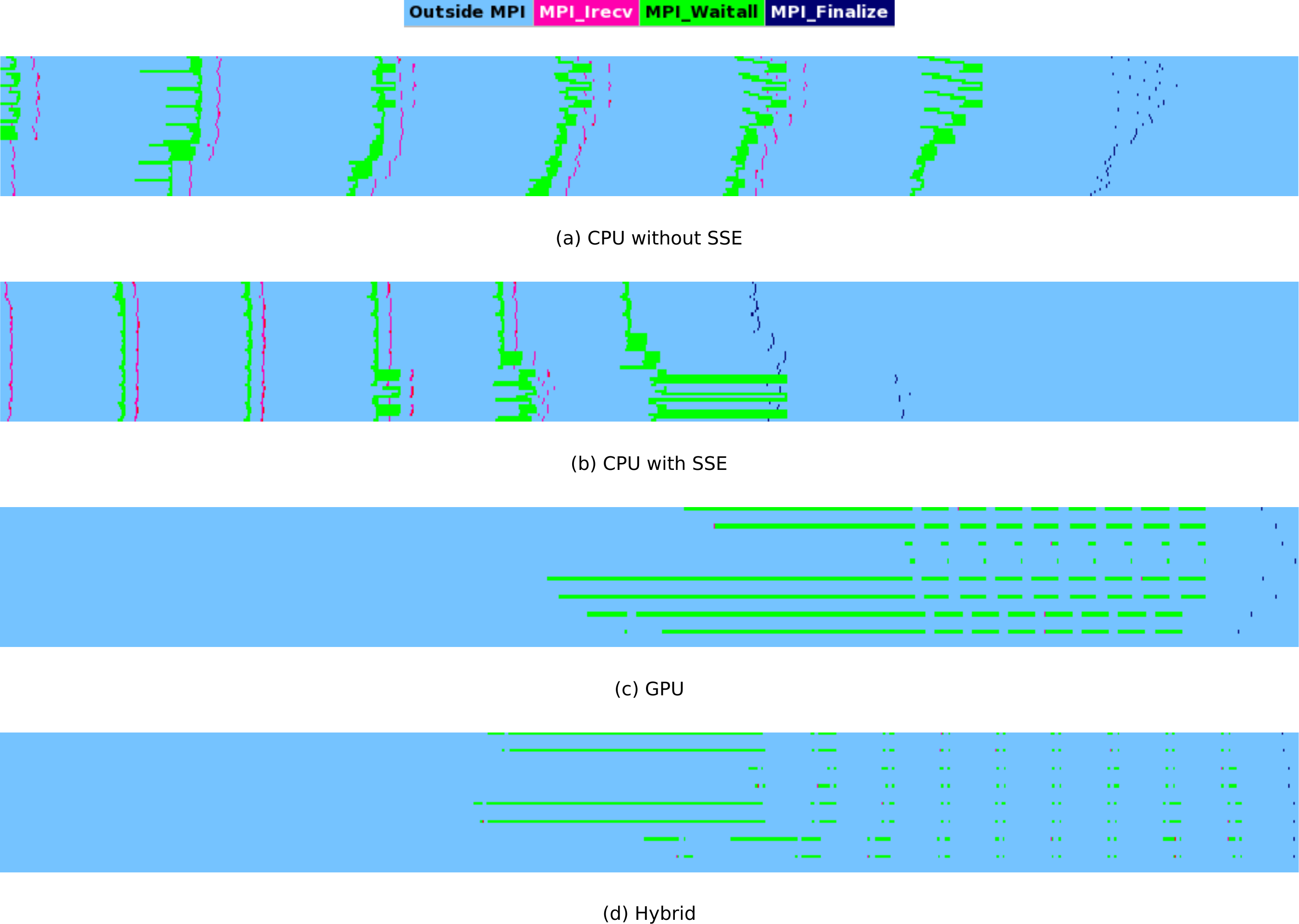}
\caption{MPI traces on four nodes}
\label{fig:paraver}
\end{figure*}

\begin{figure}[ht!]
\centering
\includegraphics[width=\columnwidth]{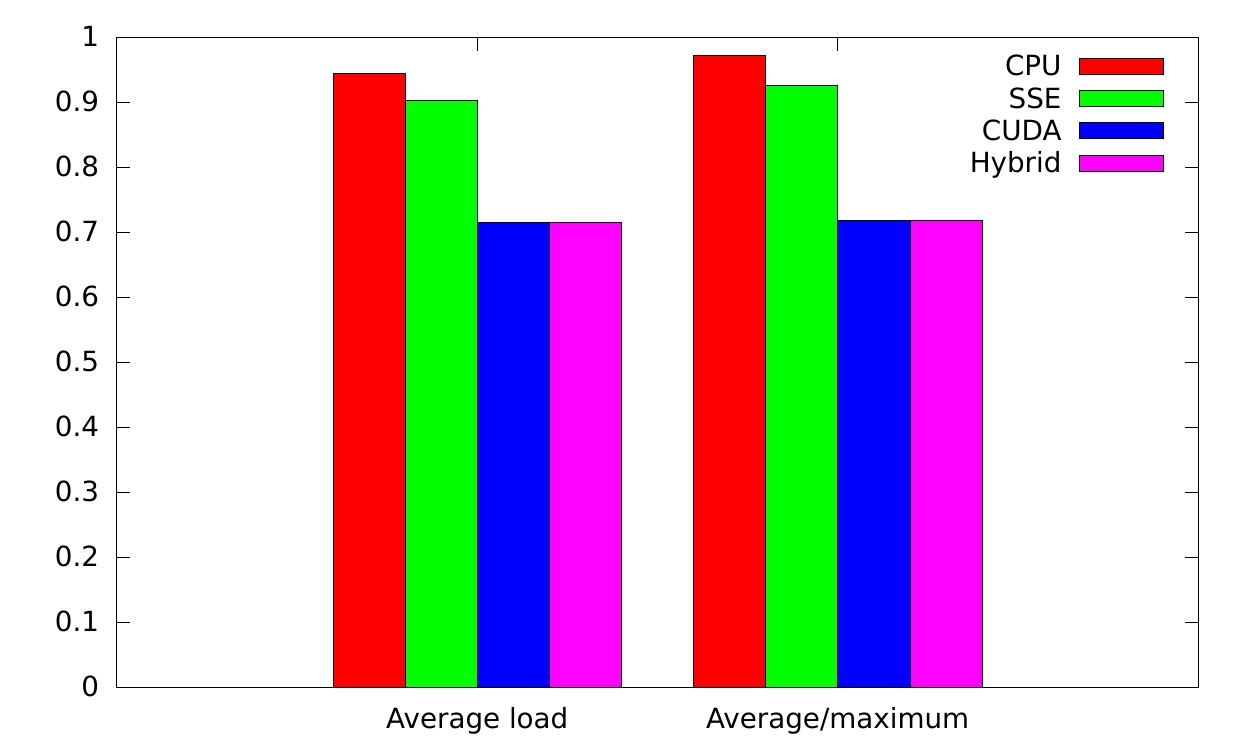}
\caption{Load balance on four nodes}
\label{fig:loadbalance}
\end{figure}

\subsection{Benchmark results}
The benchmarks ran ten iterations on increasing cluster sizes, using a synthetic state in a square domain of different lengths $L$, where $L=24576$, $49152$, and $98304$ in single precision, and $L=17472$, $34944$, $69888$ in double precision. 
Note that since this is a two dimensional quantum system, the dimension of the corresponding Hilbert space is $L^2$, and the Hamiltonian matrices would have size $L^2 \times L^2$. The dimensions were chosen so as to fill the device memory on cluster sizes one, four, and sixteen nodes. GPUs have a better a performance when the load is higher, whereas CPUs are less sensitive to the load. Choosing the matrix dimensions to fit the GPUs in certain configurations shows how the overall GPU performance decays as the cluster size increases, and it is also fair with respect to the CPU kernels due to their insensitivity to the load. The timing results are plotted in Figures \ref{fig:exec2}, \ref{fig:exec4}, and \ref{fig:exec9}. The results show only the time taken in the main loop of the evolution (as depicted in Figure \ref{overview}), as the initalization takes considerably different amounts of time with different types of kernels.

The CPU kernels show an almost linear scaling: the execution time is divided by approximately two as the cluster size is doubled. Communication overhead increases with the cluster size, so eventually the advantage of SSE optimization vanishes with large clusters (see also the next section on parallel efficiency).

The GPU kernel has a more interesting scaling pattern. When the device memory is loaded to at least 50 \%, the scaling is close to linear, just as in the case of CPU kernels. Then the execution time of individual GPUs remains almost constant, the curve flattens out, and there is little benefit to gain by this kernel in large clusters. 

The hybrid kernel trails the GPU kernel in cases where the problem would fit the GPU memory, with the execution time being marginally longer. The real advantage is in cases where the device memory is insufficient. In such cases, the speedup can be close to 2x in double precision compared to the CPU kernels.

\subsection{Parallel efficiency}
Analysing the MPI traces reveals important information about the communication patterns (Figure \ref{fig:paraver}). We restrict our attention to four nodes alone, other cluster sizes show similar patterns. The CPU kernels communicate in an almost identical pattern irrespective of the SSE optimization. Both of them are quite close to being optimal. Without SSE optimization, the average load is 94 \%, so on average, the communication overhead is approximately 6 \% (Figure \ref{fig:loadbalance}). The variation of load is small, the average/maximum ratio is 0.97. The SSE-optimized kernel has slightly worse indicators, with an average load of 90 \%. This indicates that as the computation gets more efficient, the communication becomes a bottleneck.

The GPU kernel apparently has a very different pattern. Only a fraction of the processes do any kind of work, the ones that are associated with a GPU. The plot of the MPI trace does not show the time spent in the CUDA kernel, since the launch is asynchronous with streams. Having no computational load, the CPU spends most of its time communicating, resulting in an average load of barely 71 \%, and an average/maximum ratio of 0.72, meaning that there is little variation across the processes. 

The trace of the hybrid kernel is surprising, as it looks similar to the GPU trace. While there are considerably more threads, the ones that are associated with GPU operations follow the same pattern as above, resulting in long waits. The rest of the threads, however, overlap communication extremely efficiently. The overall load balance and parallel efficiency are very similar to the GPU case.

\section{Conclusions and Future Work}\label{future}

In this paper we have shown a distributed variant of the Trotter-Suzuki algorithm based on efficient kernel implementations. We have improved the single-node efficiency of CPU kernels by replacing OpenMP directives by explicit MPI parallelization, and the GPU kernel by using streams. We have shown that our algorithm scales almost ideally, and that our hybrid kernel  is efficient for calculating the evolution of large systems in smaller clusters.

The implementation can be improved in different ways. The current breakdown of tasks is entirely manual and hard-coded: halo calculation, halo communication, and calculation of internal cells, the latter which might be split between CPU and GPU resources. When we regard the computations alone, there is a well-defined task, the calculation of a block. The block size is different on the CPU and the GPU. The former is larger, but it is an integer multiple of the GPU block size, so we can use the CPU block size as the unit of calculation. Our implementation is a typical thread-parallel approach, where heavy processes perform the work. Task-based parallelism is another approach, and the task in our case would be the calculation of a block. OmpSs is a variant of OpenMP which takes this approach to parallelism \cite{bueno2011productive}. It handles heterogeneous hardware that includes GPUs and CPUs, and takes care of the memory copies based on explicitly expressed data dependencies. Asynchronous communication can also be defined as a task. Hence theoretically it is possible to have a hybrid approach that is not hard-coded, but the distribution of halo calculation and internal cell calculation is decided by the OmpSs runtime. This also means that part of the halo might be calculated by the GPU, and it should also be easier to work on a cluster where some nodes have GPUs and others do not. To achieve this flexibility, we are working on an OmpSs version of our implementation.

Another clear direction is to extend our implementation to three dimensional systems. The variety of possible decomposition strategies in this case is large, and a flexible implementation would be very useful to test out the performance of the different choices. The extension to three dimensions is also motivated by recent theoretical and experimental developments in ultracold atomic gases \cite{lewenstein2007ultracold}, which could not be simulated in a single computer with enough precision.



\section{Acknowledgment}
This work was carried out while P. W. was visiting the Department of Computer Applications in Science \& Engineering at the Barcelona Supercomputing Center, funded by the ``Access to BSC Facilities" project of the HPC-Europe2 programme (contract no. 228398).

\bibliographystyle{elsarticle-harv}
\bibliography{bibliography}

\end{document}